\documentclass{moriond}
\usepackage{cite}
\usepackage{amsmath}
\usepackage{amssymb}

\def\be{\begin{equation}}
\def\ee{\end{equation}}
\def\bea{\begin{eqnarray}}
\def\eea{\end{eqnarray}}

\newcommand{\Photo}{}

\begin{document}
\vspace*{4cm}
\title{Uncovering stochastic gravitational-wave backgrounds with LISA}

\author{Quentin Baghi, Marc Besan\c{c}on}
\address{CEA Paris-Saclay University, Irfu/DPhP, \\ Bat. 141, 91191 Gif sur Yvette Cedex, France}

\author{Nikolaos Karnesis}
\address{Department of Physics, Aristotle University of Thessaloniki, \\Thessaloniki 54124, Greece}

\author{Jean-Baptiste Bayle}
\address{University of Glasgow, \\ Glasgow G12 8QQ, United Kingdom}

\author{Henri Inchauspé}
\address{Institut für Theoretische Physik, Universität Heidelberg, \\ Philosophenweg 16, 69120 Heidelberg, Germany}

\maketitle
\abstracts{
Finding a stochastic gravitational-wave background (SGWB) of astrophysical or primordial origin is one of the quests of current and future gravitational-wave observatories. While detector networks such as LIGO-Virgo-Kagra or pulsar timing arrays can use cross-correlations to tell instrumental noise and SGWB apart, LISA is likely to be the only flying detector of its kind in 2035. This particularity poses a challenge for data analysis. To tackle it, we present a strategy based on Bayesian model selection. We use a flexible noise power spectral density~(PSD) model and the knowledge of noise and signal transfer functions to allow for SGWBs detection when the noise PSD is unknown. With this technique, we then probe the parameter space accessible by LISA for power-law SGWB shapes.}

\section{Introduction}

One of the eight science objectives of the future space-based gravitational-wave (GW) detector LISA is directly detecting a stochastic gravitational-wave background (SGWB). Such a signal could be either of cosmological or astrophysical origin. A cosmological background could be produced in the primordial Universe through first-order phase transitions, cosmic strings or specific inflation scenarios~\cite{Caprini2018mtu, Bartolo2016ami}. Its discovery would bring a unique insight into the conditions prevailing in the early Universe and help understand TeV-scale particle physics in regions of space-time beyond the last scattering surface, i.e., which are inaccessible to us through light. Besides, unveiling an astrophysical GW foreground would give us valuable details on the population of compact objects in the Universe, like the stellar-mass black hole binaries the LIGO-Virgo network has been observing.

Detecting a SGWB with LISA is a tremendous challenge precisely because of the difference in LISA data content compared with current terrestrial detectors. First, signals from bright, compact binary sources will dominate the noise in part of its frequency band. As SGWBs are likely to be faint and necessitate long observation times, detecting them will require an accurate and complete characterisation of all resolvable GW sources in the data. These include tens of thousands of compact stellar binaries, possibly hundreds of supermassive black hole coalescences, and as many extreme-mass ratio inspirals. The hunt for a SGWB can only be done as one block of this high-dimensional inference. Second, the analysis must rely on accurately characterising the instrumental noise. One cannot use the cross-correlation methods developed for the ground-based network unless several millihertz detectors fly simultaneously. These methods assume that correlations among distant detectors are sourced by GWs only~(see \cite{Christensen_2019,romano_detection_2017} for reviews), neglecting other correlated perturbations. However, this assumption may no longer be valid in future observing runs~\cite{Janssens2023}. In the LISA case, one must rely on the knowledge of the detector response to the noise and signal to tell them apart, using prior information on their spectral features.

Significant efforts have been made to develop methods to separate noise and GW stochastic components, as well as different types of SGWBs~(see, e.g., \cite{Cornish2001bb,Caprini2019pxz,Pieroni2020rob,Flauger2020qyi,Boileau2020rpg,Karnesis2021tsh}). These searches usually rely on a parameterised noise spectrum, where the amplitudes of two noise components with fixed spectral shapes must be estimated. To increase the model flexibility, we have investigated an approach to distinguish SGWB templates from an instrumental noise of a completely unknown spectral shape. We also used realistic simulations of the signal measurement in the time domain, featuring time-varying armlengths and second-generation time-delay interferometry (TDI)~\cite{tinto_time-delay_2020}. We report our result in this communication to make a proof of principle that will be used to build an analysis with increased robustness against noise modelling errors.
% In contrast to previous search methods where the \gls{lisa} instrumental noise was parametrized with a fixed and known spectral shape, we investigate in this paper an approach to distinguish a simple \gls{sgwb} signal from the instrumental noise assuming that the single-link interferometric noises have an arbitrary and unknown spectrum

\section{Measurement model}\label{sec:model}

LISA forms a network of laser interferometers linking 6 test masses onboard 3 satellites distant from each other by 2.5 million kilometres~\cite{Bayle2023}. The $N_c$ delivered interferometer measurements can be gathered in a multivariate time series $\mathbf{y}$, which includes $N$ time samples and $N_c$ channels. These time series can be broken down into three components: laser frequency noise $\mathbf{p}$, other instrumental noises $\mathbf{n}$ (also called secondary noises) and GW signal $\mathbf{y}_{\mathrm{GW}}$ which is only significant in the long-arm interferometer measurements. We can write:
\begin{eqnarray}
	\mathbf{y} = \mathbf{y}_{\mathrm{GW}} +  \mathbf{n} + \mathbf{p},
\end{eqnarray}
where all the vectors are columns of size $N N_c$. Due to the large distance that the long-arm beam is travelling, the laser frequency noise overwhelmingly dominates over secondary noises and GW signals: $\lVert \mathbf{p} \rVert \gg \lVert\mathbf{n}\rVert,  \lVert\mathbf{y}_{\mathrm{GW}} \rVert$. Thus, the TDI post-processing technique is applied on the ground once the data is received from the spacecraft to mitigate the laser frequency noise. This operation can be modelled by a $3N \times NN_c$ matrix $\mathbf{M}$ designed to cancel $\mathbf{p}$. It yields 3  TDI variables $\mathbf{d} = \left(\mathbf{X}^{\intercal}, \mathbf{Y}^{\intercal}, \mathbf{Z}^{\intercal} \right)^{\intercal}$ which are given by
\begin{eqnarray}
	\mathbf{d} = \mathbf{M} \mathbf{y} \approx \mathbf{M}  \left(\mathbf{y}_{\mathrm{GW}} + \mathbf{n}  \right),
\end{eqnarray}
where both secondary noises and GW signals are transformed the same way through TDI. In principle, the 3 TDI variables contain all the scientific information. 

Now we assume that i) the GW signal is purely stochastic (ignoring all bright, resolvable sources), ii) the noises are zero-mean, stationary and Gaussian and iii) the SGWB is isotropic and stationary. Then, the frequency bins of the Fourier-transformed data $\mathbf{\tilde{d}}$ are approximately uncorrelated and can be described by a zero-mean multivariate Gaussian process of dimension~3 with covariance $\mathbf{C}_{d}$ given by 
\begin{eqnarray}
	\mathbf{C}_{d}(f) = \mathbf{M}(f) \left(\mathbf{C}_{\mathrm{GW}}(f)  + \mathbf{C}_{n}(f) \right) \mathbf{M}^{\dagger}(f),
\end{eqnarray}
for each frequency bin $f$, where $\mathbf{C}_{\mathrm{GW}}$ and $\mathbf{C}_{n}$ are the $N_c \times N_c$ covariances of the single-link signal and noise, respectively. The problem we wish to tackle is how to characterize $\mathbf{C}_{\mathrm{GW}}$ when we do not have accurate a-priori information on $ \mathbf{C}_{n}$.

\section{Distinguishing between signal and noise}
One can approach the problem of detecting SGWBs through a non-parametric method for the signal, ensuring agnosticism towards its generation mechanism, while using a parametric model for the noise~\cite{Caprini2019pxz}. In the present work, we tackle the problem oppositely: we use a parameterised template for the SGWB and a non-parametric model for the noise. This methodology is motivated by the discrepancy of the observed acceleration noise in the LISA Pathfinder mission compared to pre-flight models~\cite{LPF2016}. Although the mission was also a success, this feedback teaches us to be cautious about the accuracy of physical noise models.

The underlying idea of the method is to use both a parametrised SGWB power spectral density~(PSD) and a flexible noise model that can describe any spectral shape. The knowledge of the covariance structure of the two components (driven by how they inter the interferometric measurements) will help disentangle them. One can express the signal covariance as the product of a frequency-dependent correlation matrix $\mathbf{R}(f)$ by the PSD of the SGWB~\cite{Baghi_2023}:
\begin{eqnarray}
	\mathbf{C}_{\mathrm{GW}}(f) = \mathbf{R}(f) S_{h}(f).
\end{eqnarray}
The SGWB spectrum $S_{\mathrm{GW}}(f)$ can be described by any parameterized model.

In general, the noise covariance structure is more complicated. The $N_c$ interferometric measurements can be correlated (primarily through test-mass noise) and exhibit different noise levels or spectral shapes. Consequently, the $N_c \times N_c$ spectrum matrix $\mathbf{C}_{n}(f)$ may have a very general structure, where all diagonal entries are described by different PSDs and all off-diagonal entries by different cross-spectral densities (CSDs):
\begin{eqnarray}
{C}_{n}(f)[i, j] = S_{ij}(f)\text{   } \forall i, j \in \left[1, \, N_c\right]^2.
\end{eqnarray}
In our setup, we assume that all these functions are unknown. However, to simplify the problem and provide a first proof-of-concept, we make two crucial assumptions: we assume that all channels have the same PSDs and that all the cross-terms are zero: $S_{ij}(f) \approx S_{n}(f) \delta_{ij}$.
We are left with a unique noise PSD function $S_{n}(f)$ to fit for which model using a set of cubic B-spline functions
\begin{eqnarray}
	\label{eq:splines}
	\log S_n(f)=\sum_{i=1}^{Q+1} a_i B_{i, 3}(\xi, f),
\end{eqnarray}
where the spline coefficients $a_i$ and knots $\xi_j$ need to be estimated from the data. The data can then analysed using the Whittle likelihood
\begin{eqnarray}
	\label{eq:likelihood}
	p(\mathbf{\tilde{d}} | \boldsymbol{\theta}) = \prod_{k=1}^{N_f} \pi^{-3} \lvert \mathbf{C}_{d}(f_k) \rvert^{-1/2} \exp\left\{ - \mathbf{\tilde{d}}^{\dagger}(f_k) \mathbf{C}_{d}(f_k)^{-1} \mathbf{\tilde{d}}(f_k) \right\},
\end{eqnarray}
where $N_f$ frequency bins $f_k$ are analysed, and $\boldsymbol{\theta}$ includes the spline parameters in Eq.~\eqref{eq:splines} and the signal parameters describing $S_{h}$. Note that approximating the covariance using a reduced frequency grid can be useful to decrease the computational cost, see~\cite{Baghi_2023}. 

\section{Application to simulated datasets}

We apply the method described above to simulated LISA data streams. We perform the simulation in the time domain to account for all possible artefacts related to the transformation of finite time series into the frequency domain. We generate the interferometric data $\mathbf{y}$ with the \texttt{LISA GW Response}~\cite{lisagwresponse} code which approximates this sky as incoherent stochastic point sources which are evenly spread on the celestial sphere using \texttt{HEALPix}~\cite{Zonca2019}. Then, we obtain the TDI time series $\mathbf{b}$ by processing \texttt{LISA GW Response}'s outputs with the \texttt{PyTDI}~\cite{pytdi} software, which combines the measurements and applies adequate delays based on Lagrange fractional delay filters to form the 3 Michelson variables. The noise component is then added to the signal by generating independent Gaussian realisations in the frequency domain weighted by the PSD $S_{n}(f)$ and returning to the time domain.

\begin{figure}[!h]
	\centering
	\includegraphics[width=0.6\textwidth, trim={0.6cm 0.8cm 1cm 0.6cm}, clip]{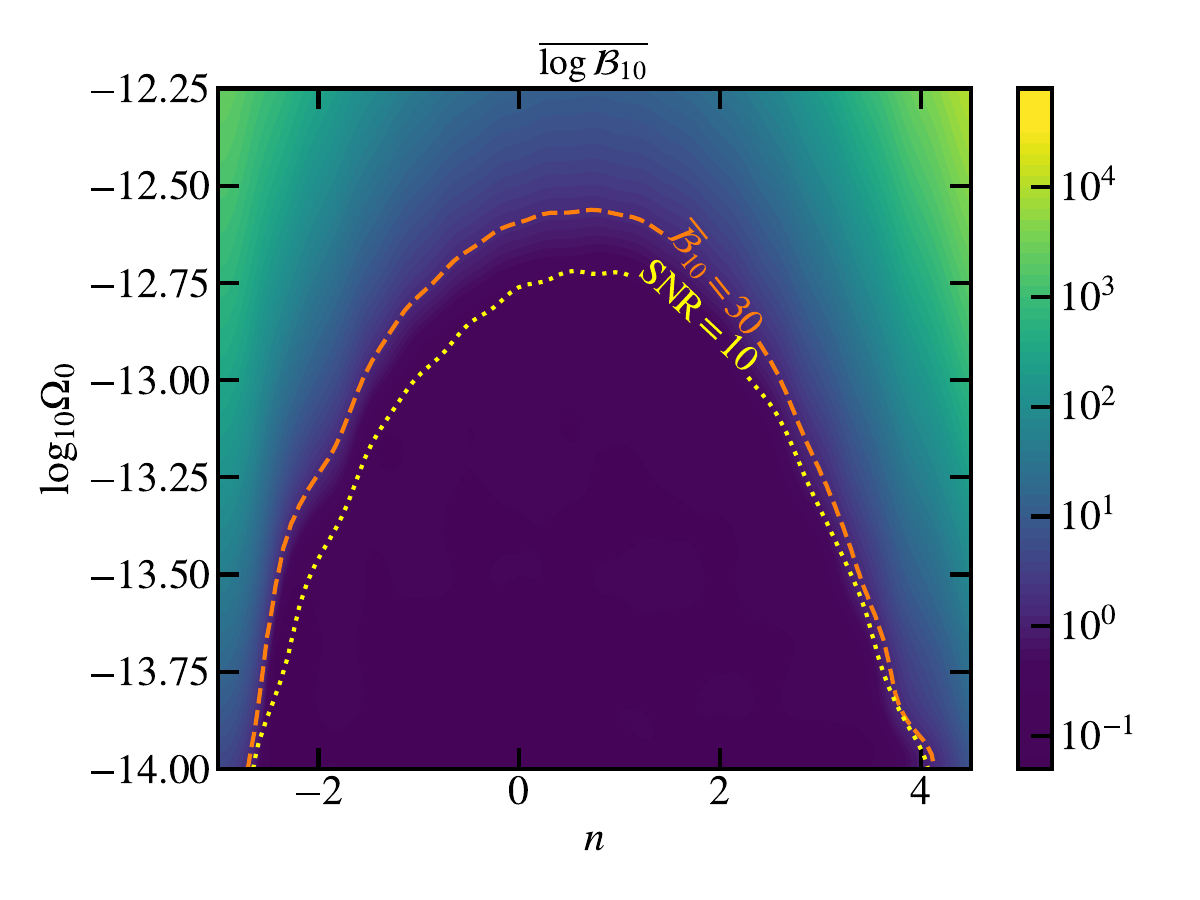}
	\caption{Contours of log-Bayes factors comparing a model with ($H_1$) and without ($H_0$) SGWB as a function of amplitude $\Omega_0$ and power index $n$ for a power-law model. The orange dashed line indicates the  detection threshold.}
	\label{fig:bayes-factor}
\end{figure}

The injected signal has an energy density which follows a power-law $\Omega_{\mathrm{GW}}(f) = \Omega_{0} (f / f_0)^{n}$, so that the SGWB PSD is $S_{h}(f) = \Omega_{\mathrm{GW}}(f) 3H_{0}^2 / (4 \pi^2 f^3)$. Then, we can analyse the data based on likelihood~\eqref{eq:likelihood} under two hypotheses: $H_0$, in which the data only contain noise and $H_1$, in which the data contains noise and signal. We use Bayesian model comparison to explore the parameter space for which detection is possible, using the Bayes factor as a criterium. We obtain the result shown in Fig.~\ref{fig:bayes-factor}.
All parameter pairs $(\Omega_0,\, n)$ that lie above the dashed orange curve correspond to GW signals that should be detectable with high confidence, despite the flexibility of the non-parametric modelling we propose. Note that such an analysis can be reproduced for any SGWB template.

\section{Going beyond}

The reported study lays the grounds for robust detection of SGWBs when a-priori information on noise is unreliable. However, many investigations and improvements must be included to handle realistic LISA data. First, the diagonal approximation of the single-link covariance matrix $\mathbf{C}_{n}$ needs to be relaxed to include correlations between interferometric measurements and unequal PSDs. Second, the degeneracy between signal and noise should be studied in a fully non-parametric setting. Third, contamination from resolvable GW sources in the data should be investigated, together with the presence of non-stationary foregrounds. Furthermore, using state-of-the-art instrumental noise simulations would help improve and demonstrate LISA's ability to discover the backgrounds of GWs.

\section*{Acknowledgments}
%\vspace{-0.2cm}
The authors thank the LISA Simulation Expert Group for all simulation-related activities. They would like to personally thank J. Veitch for their insightful feedbacks. J.-B.B. gratefully acknowledges support from UK Space Agency (grant ST/X002136/1). N.K. acknowledges support from the Gr-PRODEX 2019 funding program (PEA 4000132310). %Some of the results in this paper have been derived using the healpy and HEALPix package.

%\section*{Appendix}
%
% We can insert an appendix here and place equations so that they are
%given numbers such as Eq.~\ref{eq:app}.
%\be
%x = y.
%\label{eq:app}
%\ee
%
%\section*{References}

\bibliography{references}
%\begin{thebibliography}{99}
%\bibitem{ja}C Jarlskog in {\em CP Violation}, ed. C Jarlskog
%(World Scientific, Singapore, 1988).
%
%\bibitem{ma}L. Maiani, \Journal{\PLB}{62}{183}{1976}.
%
%\bibitem{bu}J.D. Bjorken and I. Dunietz, \Journal{\PRD}{36}{2109}{1987}.
%
%\bibitem{bd}C.D. Buchanan {\it et al}, \Journal{\PRD}{45}{4088}{1992}.

%\end{thebibliography}

\end{document}